\documentclass[10pt]{article}
\usepackage[margin=1.0in]{geometry}
\usepackage{setspace}
\usepackage{indentfirst}
\usepackage{caption}
\usepackage{subcaption}
\usepackage{amsmath}
\usepackage{amssymb}
\newcommand{\beq}{\begin{equation}}
\newcommand{\eeq}{\end{equation}}

\usepackage{graphicx}
\graphicspath{{./JPG_Images/}}
\usepackage{url}
\title{
A comparison of heroin epidemic models}

\author{
Nicholas A. Battista  \\
Center for Applied Mathematics \\
School of Mathematical Sciences, Rochester Institute of Technology,\\
85 Lomb Memorial Drive, Rochester, NY 14623-5603, USA \\ $ $ \\
May 21, 2009}

\date{}
\begin{document}
\pagestyle{plain}

   \maketitle
   \setlength{\parindent}{0pt}
   \begin{abstract}

The use of illicit drugs has been on the rise in United States. It is very detrimental on society, as fatal overdose is the fourth leading cause of death in the United States, which is about the same as motor vehicle crashes. Of all illicit drugs, one drug that has an severe adverse effect on a community as a whole is heroin. This paper will discuss two mathematical models- the White and Comiskey model and a newly introduced model proposed by the author, describing heroin use within a fixed community. We will show the existence of stable equilibrium from both models, suggesting both a situation where heroin use is eradicated and one where it remains an endemic.

    \vspace{.4in}
     \end{abstract}

%
%

\section{Introduction}

The abuse of illicit drugs is detrimental on society. Besides the dangers and risks in obtaining and using the drug, the drug users also pose imminent health threats to the rest of a community, since they are more susceptible to disease to which they can then transmit to others from their immune system being damaged from drug abuse. One such illicit drug that has severe effects in society is heroin.

The number of heroin first users has nearly tripled between 2005 and 2006. \cite{NIDA:2008} Furthermore data collected from the 2008 National Survey on Drug Use and Health (NSDUH) suggests that there are approximately 1,200,000 million Americans who use heroin \emph{recreationally}, and about 200,000 Americans, who are regular users, in 2007 \cite{NSDUH:2008}. One critical aspect of heroin use is thus far its highest numbers of users fall into a category of being 26 years old or older, where the average age of first use is approximately 20.7 years \cite{NIDA:2007}. This statistic is critical in describing how heroin abuse has been contained, since it is not being used commonly in high schools, where kids are more susceptible to peer pressure and hence falling into drug abuse at earlier ages.

Heroin users commonly face many serious, and possibly detrimental health conditions, from pulmonary complications to spontaneous abortion. Habitual heroin users experience poor health resulting from the drug weakening their immune system. As mentioned before, this makes them more susceptible to falling ill and transmitting a disease through society. Also, simply the way the drug is taken is dangerous, since it is most of the times injected through needles. This increases the abusers risk of infectious diseases, including hepatitis and the HIV/AIDS virus \cite{NIDA:2007}.

The rehabilitation and treatment of heroin users is very costly and a major burden on society \cite{Mulone:2009}. Medical treatment is usually integrated with other supportive services, in hopes of having the patients return to lead stable and productive lives. However, it is very difficult to wane abusers off the drug since heroin is dangerously addictive. Only $20\%$ of heroin users enter into treatment one year from their first use. Moreover, takes roughly 3 years since their first use of heroin for $50\%$ of users to seek treatment \cite{Frisher:2006}. Unfortunately, the number of years since first usage and entering treatment has increased since the 1970s \cite{Hunt:1974}, \cite{Frisher:2006}. Since detoxification is the first step in rehabilitation, drug aids have been introduced to help users experience less severe withdrawal symptoms. Such drugs are clonidine and buprenorphine. Unfortunately, since those drugs were created to by medical professionals to stimulate a lot of the same brain's receptor's as heroin would, the detox patients often become dependent on them while in treatment. \cite{NIDA:2007}

In order to find the most effective of way of treating the drug, as well as finding how a drug epidemic might start and spread through a community, mathematical modeling will be used. Mathematical modeling has been extensively used in population ecology \cite{Allen:2007}, and has recently started to be applied to drug epidemic models \cite{Mackintosh:1979}, \cite{Rossi:2002},\cite{Winkler:2002}, \cite{Rossi:2004}, \cite{White:2007}, This paper will discuss two mathematical models describing the spread of heroin, the White and Comiskey model \cite{White:2007} and a newly proposed model. Both models are SIR based models and show the existence of an deal equilibrium, where drug use is eradicated, and the existence of endemic steady-states, where drug use remains in society.

%
%

\section{White and Comiskey Model}

The White and Comiskey heroin epidemic model assumes that there are only three classes of people- susceptible, S, drug users, $U_{1}$, and drug users undergoing treatment, $U_{2}$ \cite{White:2007}. The susceptible class is composed of people who have never used the drug before. The drug users are the people who begin using the drug and the drug users undergoing treatment are those people who go into medical treatment to battle their addiction.

There are two key parameters in the model, $\beta_{1}$ and $\beta_{3}$. $\beta_{1}$ is the probability of becoming a drug user per unit time, and $\beta_{3}$ is the probability that a drug user in treatment relapses back into untreated drug abuse. When considering the drug to be heroin, from data collected in Dublin, it was found that $\beta_{1}<\beta_{3}$. Hence they find that $\beta_{1} \sim 0.02$ and $\beta_{3}\sim 0.8$, suggesting that the probability of relapse is much higher than the probability of falling into drug use to begin with. It is important to note that these numbers are not expected to vary much in other regions.

%
%

\subsection{Model}

The box flow diagram for this model is depicted in Figure(1).\\

\begin{figure}[!h]
  \centering
  \includegraphics[scale=0.3]{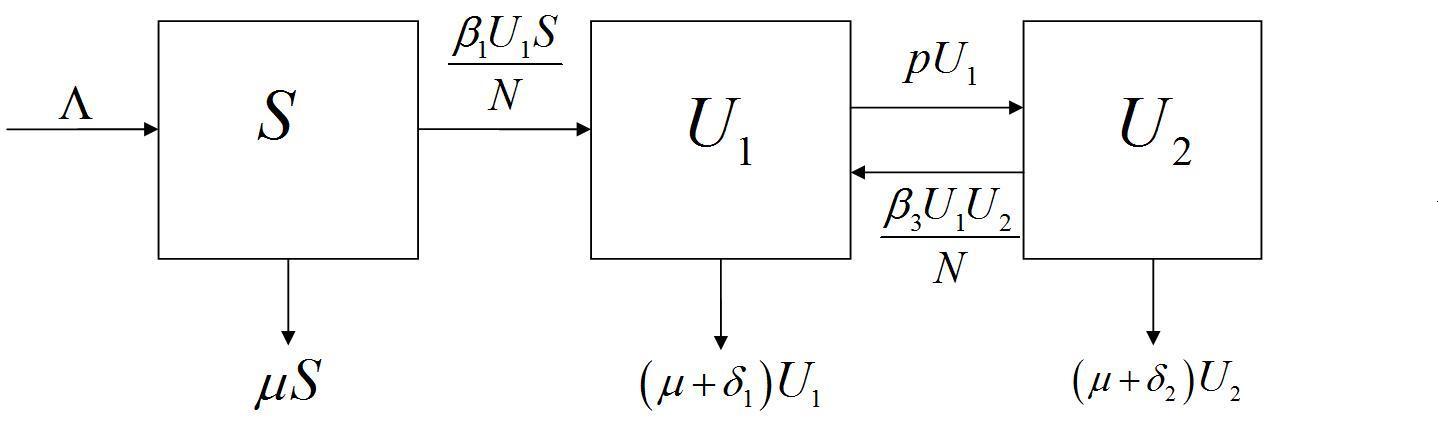}
 \footnotesize{\caption{The population flow diagram for the White and Comiskey model. The susceptible class can either become drug users, or leave the system, by either passing away or refusing to begin use of the drug. The drug user class can leave by passing away or quitting use, or entering treatment. The drug users in treatment can either relapse back into drug use, pass away, or quit forever.}}
  \label{WhiteAndComiskeyBoxes}
\end{figure}

The flow can be described as follows. The susceptible class can either "smarten-up" (or die) and leave, or they can become drug users.
The drug users can either continue using the drug, "smarten-up" or die, or enter medical treatment.
The drug users undergoing treatment can then either relapse back into the drug or "smarten-up" (or die).
 This model is described as a dynamical system composed of 3 coupled, non-linear ordinary differential equations as follows,
    \begin{eqnarray}
    \label{Model1S} \frac{dS}{dt} &=& \Lambda - \frac{\beta_{1} U_{1} S}{N} - \mu S \vspace{1in} \\
    \nonumber \\
     \frac{dU_{1}}{dt} &=& \frac{\beta_{1} U_{1} S}{N} - pU_{1} + \frac{\beta_{3} U_{1} U_{2}}{N} - (\mu + \delta_{1}) U_{1}\nonumber \\
     \label{Model1U1} \\
    \label{Model1U2} \frac{dU_{2}}{dt} &=& pU_{1} -  \frac{\beta_{3} U_{1} U_{2}}{N} -  (\mu + \delta_{1}) U_{2}.
    \end{eqnarray}

The parameters of the model are defined as follows:
    {\footnotesize{
    \begin{itemize}
    \item $\Lambda:$ Number of individuals entering susceptible population
    \item $\mu:$ Natural death rate of general population
    \item $\delta_{1}:$ Enhanced removal rate/death rate for drug users
    \item $\delta_{2}:$ Enhanced removal rate/death rate for drug users seeking treatment
    \item $p:$ Proportion of drug users who enter treatment per unit time
    \item $N:$ The total population ($N=S+U_{1}+U_{2}$).
    \end{itemize}}}

Since the model assumes a fixed population, we note that $\Lambda = \mu S + (\mu+\delta_{1})U_{1} + (\mu+\delta_{2})U_{2}.$ We also introduce the following non-dimensionalized variables,
\begin{equation}
\label{WhiteAndComiskeyNonDimVar} s=\frac{S}{N}, \ \ u_{1} = \frac{U_{1}}{N}, \ \ u_{2} = \frac{U_{2}}{N}.
\end{equation}

Note that $s+u_{1}+u_{2}=1$. Using the above value of $\Lambda$ and Eq.(\ref{WhiteAndComiskeyNonDimVar}) we can eliminate a variable in the model. When eliminating $u_{2}$ via the relation that $u_{2}=1-s-u_{1}$, the model becomes
%
\begin{eqnarray}
\label{Model2S} \frac{ds}{dt} &=& \mu + \delta_{2} + (\delta_{1}-\delta_{2}) u_{1} - (\mu+\delta_{2}) s- \beta_{1} u_{1} s,\ \ \ \ \ \  \\
\nonumber \\
\label{Model2u1} \frac{du_{1}}{dt} &=& (\beta_{3}-p-\mu-\delta_{1}) u_{1} + (\beta_{1}-\beta_{3}) u_{1} s - \beta_{3} u_{1}^{2}.
\end{eqnarray}
%
From the above system, we will solve for the equilibrium points by setting Eqs.(\ref{Model2S}) and (\ref{Model2u1}) equal to zero. Upon doing so we find that there will be a trivial equilibria, $(s^{*},u_{1}^{*},u_{2}^{*}) = (1,0,0)$, which we will call the \emph{drug-free\ equilibria,} and an endemic equilibrium point, which depends on values of the parameters.

%
%

\subsection{Stability Analysis of Equilibria}

To study the stability of the equilibrium points we first construct the Jacobian matrix, defined as the matrix of first partial derivatives of each equation, $J_{kl} = \frac{\partial f_{k}}{\partial x_{l}}.$

The general Jacobian for the White and Comiskey model is given below,
\begin{equation}
\label{WhiteAndComiskeyJac} J(s^{*},u_{1}^{*}) = \left( \begin{array}{cc}
-\beta_{1}u_{1}-(\mu+\delta_{2}) & \delta_{1} - \delta_{2} -\beta_{1}s\\
$ $ & $ $\\
(\beta_{1}-\beta_{2}) u_{1} & \beta_{3}-p-\mu-\delta_{1}+(\beta_{1}-\beta_{3})s - 2\beta_{3} u_{1}\\
\end{array} \right)
\end{equation}
%
The sign of the eigenvalues of the above matrix at each equilibrium point will determine the stability of the system. If both eigenvalues are negative (or have negative real part), then the equilibrium is stable; otherwise if an eigenvalue is positive (or has positive real part), the equilibrium will be unstable.

We will analyze the stability of both the drug-free equilibria as well as the endemic equilibria.

%
%
%
%
\subsubsection{Drug-Free Equilibria}

The resulting Jacobian matrix in this case, where $s^{*}=1$ and $u_{1}^{*}=0$ is:
\begin{equation}
\label{WhiteAndComiskeyDFJac} J(1,0) = \left( \begin{array}{cc}
	-(\mu+\delta_{2}) & \delta_{1}-\delta_{2}-\beta_{1}s \\
	$ $ & $ $\\
	0 & \beta_{1}-p-\mu-\delta_{1} \\
	\end{array}\right).
\end{equation}

Solving the eigenvalue problem, $\|\lambda I - J(1,0)\| = 0$, for this matrix, we find the eigenvalues to be:
$$\lambda = \left\{ -\mu-\delta_{2}, \beta_{1}-p-\mu-\delta_{1} \right\}.$$
Hence if $\beta_{1}<p+\mu+\delta_{1}$, the drug-free equilibria will be \emph{stable}. This situation would suggest biologically that the probability of becoming a drug-user per unit time is less than the sum of the natural death, enhanced removal rate, and fraction of the drug-users who enter medical treatment per unit time. This describes a scenario in which more people are leaving the drug-use then entering it. Otherwise if $\beta_{1} > p+\mu+\delta_{1}$, the drug-free equilibria will be unstable.

The following phase-plane depicts the case when there is stability.
\begin{figure}[!h]
  \centering
  \includegraphics[scale=0.35]{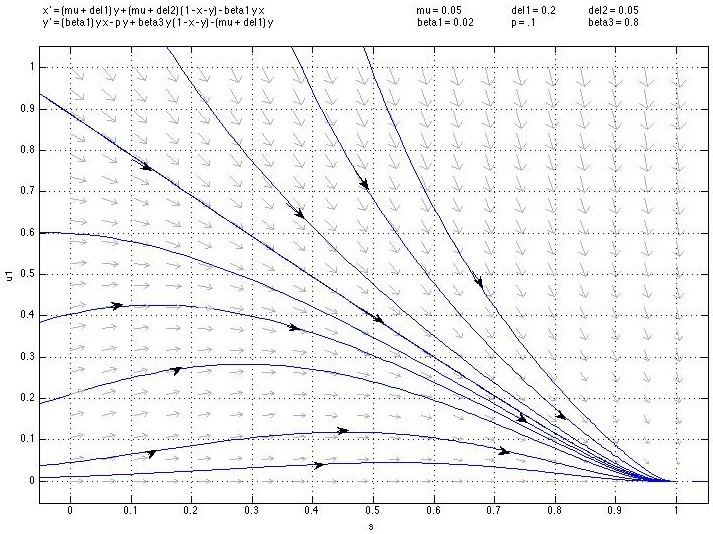}
  \footnotesize{\caption{Existence of the stable drug free equilibria in the White and Comiskey model. This phase plane plots drug users against susceptibles in a population. It is clear that this is the ideal stable equilibria, as it predicts a scenario where the total population becomes drug-free.}}
  \label{WhiteAndComiskeyDrugFree}
\end{figure}

%
%
%
%
\subsubsection{Endemic Equilibria: $\beta_{1}<\beta_{3}$}

To search for the endemic equilibria, we will assume that $\beta_{1}>p+\mu+\delta_{1}$. Assuming that $0<\beta_{1}<\beta_{3},$ which is realistic according to previous data collected, and from setting (\ref{Model2u1}) equal to zero, we get
\begin{equation}
\label{WandCEndemicS} s^{*} = -\frac{\beta_{3}u_{1}}{\beta_{3} - \beta_{1}} + \frac{(\beta_{3}-p-\mu-\delta_{1} )}{\beta_{3}-\beta_{1}}.
\end{equation}
Now substituting (\ref{WandCEndemicS}) into (\ref{Model2S}) and setting it equal to zero, we get the following quadratic equation in terms of $u_{1}$,
%

$$-\beta_{1}\beta_{3} u_{1}^{2} + \Big( (\beta_{1}-\beta_{3})(\delta_{1}-\delta_{2}) - \beta_{3}(\mu+\delta_{2}) - \beta_{1}(p+\mu+\delta_{1}-\beta_{3}) \Big) u_{1} +(\mu+\delta_{2})(\beta_{1}-p-\mu-\delta_{1})=0.$$
The positive solution of the above quadratic can be written in the following form,
\begin{equation}
\label{quadraticU1bar} \bar{u}_{1} = \frac{ - h \pm \sqrt{ h^{2} + 4\beta_{1}\beta_{3}(\mu+\delta_{2})(\beta_{1}-p-\mu-\delta_{1}   }      }{2\beta_{1}\beta_{3}},
\end{equation}\\
where $h = \beta_{1}(p+\mu+\delta_{2}-\beta_{3})+\beta_{3}(\delta_{1}+\mu).$  Substituting (\ref{quadraticU1bar}) into (\ref{WandCEndemicS}) we yield the coordinates of the endemic equilibria.

From work done by Mulone and Straughan \cite{Mulone:2009}, it is found that this equilibria is \emph{stable}. For completeness, they also found that there are no periodic orbits when considering only positive solutions. Hence when $\beta_{1}>p+\mu+\delta_{1}$ with positive initial conditions, the endemic equilibria is globally asymptotically stable.

The phase-plane in Figure(3) illustrates this.
\begin{figure}[!h]
  \centering
  \includegraphics[scale=0.4]{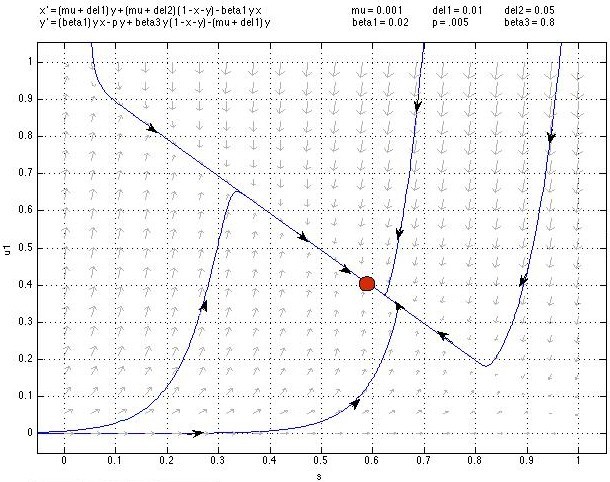}
  \footnotesize{\caption{Existence of an endemic equilibria in the White and Comiskey Model. This phase plane plots the drug-user population against the susceptible (drug-free) population. The stable point shown illustrates the a total populations that is stable with just over $60\%$ of the population drug-free and just over $40\%$ in the drug-user group. This is not an ideal equilibria, since it allows for a large percentage of the population to fall into the drug-user category.}}
  \label{WhiteAndComiskeyEndemicEquilibria}
\end{figure}

This equilibria, although stable, is not ideal for the population. As illustrated in Figure(3), it allows for a large percentage of the population to fall into the drug-user class. In this case, social programs may be implemented, which may increase the model parameter, $p$, which will decrease the drug-user population outside of treatment. 


%
%
%
%
\subsubsection{Endemic Equilibria: $\beta_{1}>\beta_{3}$}

We also investigate the case when $\beta_{1}>\beta_{3}$, that is a scenario in which a highly successful treatment procedure is implemented, allowing drug users to have far less relapses back into untreated drug abuse. Unfortunately, for heroin no such treatment has been found, yet we will analyze this case for pure theoretical and biologically motivation.

The Jacobian of the instability becomes
$$J(s^{*},u_{1}^{*}) = \left( \begin{array}{cc}
	-\beta_{1} u_{1}^{*} - (\mu+\delta_{2}) & \frac{ (\mu+\delta_{2})(s^{*}-1) }{u_{1}^{*}}\\
	$ $& $ $\\
	(\beta_{1}-\beta_{3}) u_{1}^{*} & -\beta_{3} u_{1}^{*} \\
	\end{array} \right),$$
and hence the characteristic polynomial is 

$$\lambda^{2} + \left( (\beta_{1}+\beta_{3})u_{1}^{*} + \mu +\delta_{2} \right) \lambda + (\beta_{1}u_{1}^{*} + \mu +\delta_{2}) \beta_{3}u_{1}^{*} + (1-s^{*})(\beta_{1}-\beta_{3})(\mu+\delta_{2}) = 0.$$

Because all the coefficients of the above quadratic are positive, we note that the only solutions for $\lambda$ will be when $\lambda<0$ or when $Re(\lambda) < 0$. Therefore, in this case the endemic solution will be only locally stable.

%
%

\subsection{Results and Implications}

In White and Comiskey's model, they present a dynamical system model describing the possibility of a heroin epidemic. They found that there exist only two positive steady-states- one in which is the \emph{drug-free} equilibria and another that is an endemic steady-state. \cite{White:2007}

The drug-free equilibria was found to stable under certain conditions of parameter relations. If more people are leaving drug-use then falling into it, it is found that this equilibria is stable. However, when more people begin drug abuse then those leaving it, this equilibria becomes unstable and all solutions will be attracted to the endemic steady-state. This steady-state describes a constant equilibrium between the number of susceptibles, drug users, and users in treatment, moreover it describes a situation in which the drug use still exists but remains unchanging.

There are many ways in which this model can be extended. Obviously, it could be changed to include gender differences and try to model how what role gender makes in the spread of a drug. Likewise, one could create an age based demographic model, in which they may include other parameters such as how wealthy a community is. There is also the possibility of incorporating specific parameters dealing with how much anti-drug awareness or peer pressure there is. To drastically modify the structure of the system, one could create a coupled partial differential equation (PDE) model to describe the spread of an illicit drug through a community, using a source term like a drug-lab, or big city, as its focus.

%
%

\section{New Drug Epidemic Model}

To amend the White and Comiskey model, we will assume there are more than 3 classes of people. In doing so, we assume that there are two susceptible classes, $S_{1}$ and $S_{2}$, `Light' Drug Users, L, `Heavy' Drug Users, H, and Users undergoing medical treatment, U. The susceptible-1 class is composed of people who have never used the drug before. The Light Drug users are composed of people who use the drug recreationally, or only on occasion; where as, Heavy Drug users are people who use the drug regularly and are severely addicted. The users in treatment are those people who want to give up drug use. The susceptible-2 class is made up of people who were previously users who successfully underwent treatment and saw it to completion.

We wished to amend the White and Comiskey model to incorporate multiple drug user classes. Upon doing so, we wish to create the model that describes a scenario in which the light users influence the susceptible classes to start using the drug, while the heavy users deter susceptibles from beginning drug use, since the non-users clearly observe the detrimental effect of drug use first hand. 

%
%

\subsection{Model}

The box-flow diagram depicted in Figure(4) illustrates the how a drug epidemic may spread.

\begin{figure}[!h]
  \centering
  \includegraphics[scale=0.3]{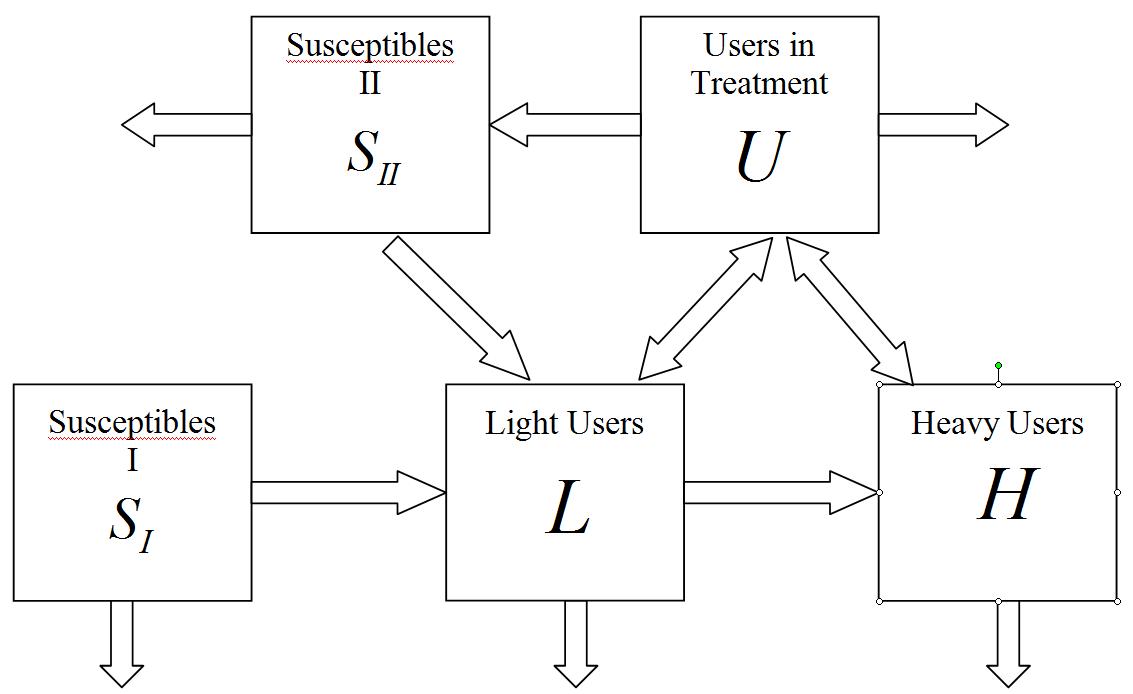}
  \footnotesize{\caption{Illustrating the population flow through the newly proposed model. There are two susceptible classes, one in which the population has never used the drug, while the other is composed of previous drug users who have successfully gone through treatment. There are also two drug user classes, one composed of users who use the drug recreational, or less frequently, while the other is composed of drug users, who use regularly. Both drug user classes can enter the users in treatment class.}}
  \label{WhiteAndComiskeyBoxes}
\end{figure}

The flow can be described as follows. The susceptibles-1 class can either become light drug users or ``smarten" up (or die).  Once a light drug user, they will either ``smarten up'' (or die), become dependent on the drug and join the heavy drug user class, or they can enter medical treatment. Once a heavy user, they can either ``smarten up" (or die) or enter medical treatment. We do not consider the case where a heavy user limits themselves in drug use to rejoin the light drug user class because we assume that the drugs are highly addictive, so an individual will not be able to limit themselves without seeking any kind of medical treatment.

Once in treatment the users can die, relapse into either the light drug user class or heavy drug user class, or they can successfully complete the treatment to join the susceptible-2 class. Then he susceptible-2 class can either become light users or they can die naturally. The reason we assume two different susceptible classes is because it is our belief that after someone uses the drug, even after they complete treatment successfully, they are still more likely to begin drug abuse again. Again, we are assuming the drug is extremely addictive.

Our model is described as a dynamical system composed of 5 coupled, non-linear ordinary differential equations as follows,

   \begin{eqnarray}
    \label{BattistaModelS1} \frac{dS_{1}}{dt} &=& \Lambda - \frac{\beta_{1} L S_{1}}{N} e^{-qH/L} - \mu S_{1} \vspace{1in} \\
    \nonumber \\
    \label{BattistaModelL}  \frac{dL}{dt} &=& \frac{\beta_{1} L S_{1}}{N} e^{-qH/L} - (\alpha_{1}+\gamma_{1} +\mu+\delta_{1})L + \sigma_{1}\frac{LU}{N} +  \frac{\beta_{2} L S_{2}}{N} e^{-qH/L} \\
    \nonumber \\
    \label{BattistaModelH} \frac{dH}{dt} &=& \alpha_{1}L - (\gamma_{2} +\mu+\delta_{2}) H + \sigma_{2}\frac{LU}{N} \\
    \nonumber \\
    \label{BattistaModelU} \frac{dU}{dt} &=& \gamma_{1}L + \gamma_{2}H - \sigma_{1}\frac{LU}{N} - \sigma_{2}\frac{LU}{N} - (\mu+\delta_{3})U - \eta U \\
    \nonumber \\
    \label{BattistaModelS2} \frac{dS_{2}}{dt} &=& \eta U -\mu\delta_{2} - \frac{\beta_{2} L S_{2}}{N} e^{-qH/L},
    \end{eqnarray}

where $\Lambda = \mu (S_{1}-S_{2}) + (\mu+\delta_{1})L + (\mu+\delta_{2})H + (\mu+\delta_{3})U,$ and describes the number of people in the general population who are susceptible to beginning drug-use. The parameters in the model are defined as follows:

    \begin{itemize}
        \item  $\Lambda:$  number of people in general population who are susceptible
        \item N: total size of population
        \item   $\beta_{1}:$ probability of becoming a drug user
        \item   $\beta_{2}:$ probability of becoming a drug user after completed treatment
        \item  q: term describing deterrent effect of heavy usage
        \item   $\mu:$ natural death rate
        \item   $\alpha_{1}:$ fraction of light users who become heavy users
        \item  $\gamma_{1}:$ fraction of light users entering treatment
	 \item  $\delta_{1}:$ removal rate for users no in treatment
        \item  $\sigma_{1}:$ probability a user in treatment relapses to light use
        \item  $\gamma_{2}:$ fraction of heavy users entering treatment
        \item  $\delta_{2}:$ removal rate of heavy users (death, ``smarten up")
        \item   $\delta_{3}:$ removal rate for those in treatment
        \item   $\eta:$  fraction of users in treatment who successfully complete it
\end{itemize}
%
In order to reduce the complexity of the model, we will use our assumption that the population is constant to write $S_{2}$ in terms of the other variables, ie- $$S_{2} = N - S_{1} - L - H - U.$$ Upon doing this, and substituting it into our model, we will reduce the dynamical system from a system of five coupled differential equations to a system of four. Furthermore, non-dimensionalizing the model using the following relations,
$$s_{1} = \frac{S_{1}}{N}, \ \ l = \frac{L}{N}, \ \ h = \frac{H}{N}, \ \ u = \frac{U}{N},$$
and then using the relation to eliminate the variable $S_{2}$, our model becomes,
   	\begin{eqnarray}
   	 	\label{BattistaNonDims1} \frac{ds_{1}}{dt} &=& \delta_{1}L + \delta_{2}h +\delta_{3}u + (1-s_{1}) \mu - \beta_{1}Ls_{1}e^{-qh/L}\\
	   	\nonumber \\
  	  	\frac{dL}{dt} &=& \beta_{1}Ls_{1}e^{-qh/L} - (\alpha_{1}+\gamma_{1}+\mu+\delta_{1})L + \sigma_{1}Lu +\ldots \nonumber \\ 			 \label{BattistaNonDimL} \ \ \ &\ & \ \ \ \ \ \ \ \ \ \ \ \ \ \ \ \ \ \ \ \ \ \ \ \ \ \ \ \ \ \ \ \ \ \ \ \ \ldots+ \beta_{2}L(1-s_{1}-L-h-u)e^{-qh/L} \\
		\nonumber \\
		\label{BattistaNonDimh} \frac{dh}{dt} &=& \alpha_{1}L  - (\gamma_{2}+\mu+\delta_{2})h +\sigma_{2}Lu\\
		\nonumber \\
		\label{BattistaNonDimu} \frac{du}{dt} &=& \gamma_{1}L + \gamma_{2}h - (\sigma_{1}+\sigma_{2})Lu - (\mu+\delta_{3}-\eta)u
   	 \end{eqnarray}
%
From the above system, we will determine the equilibria points of the system as well as their stability. We will focus on the drug-free equilibria, $(s_{1}^{*},l^{*},h^{*},u^{*}) = \left(1,0,0,0\right).$ We also note that if $l=0$ or $h=0$ then the other equals zero. So if any of the drug user classes go to zero, the other one will as well, deeming this the drug-free equilibria. Ultimately, this is the ideal equilibria. We will also present a case for endemic equilibria, where drug-abuse is still occuring, but not increasing or decreasing.

%
%

\subsection{Stability Analysis of Equilibria}
\label{stability_battista_model}

As in Section(2.2) we will construct the model's associated Jacobian matrix and then proceed to find its eigenvalues. We will do this analytically for the drug-free equilibria.  We will also show the existence of endemic steady-states; however, we will not present  its stability analysis.

\subsubsection{Stability of (1,0,0,0)}

The Jacobian in this case is found below,
$$J(1,0,0,0)=\left( \begin{array}{cccc}
	
	-\mu & \delta_{1} & \delta_{2} & \delta_{3} \\
	
	$ $ & $ $ & $ $ & $ $ \\
	
	0 & -(\alpha_{1}+\delta_{1}+\gamma_{1}+\mu) & 0 & 0 \\
	
	$ $ & $ $ & $ $ & $ $ \\
	
	0 & \alpha_{1} & -(\gamma_{2}+\mu+\delta_{2}) & 0 \\
	
	$ $ & $ $ & $ $ & $ $ \\
	
	0 & \gamma_{1} & \gamma_{2} & -(\mu+\delta_{3}-\eta) \\
	
	\end{array}\right)$$
%
We find the eigenvalues, $\lambda$, of the above matrix are
$$\lambda = \Bigg\{  -mu,-(\alpha_{1}+\delta_{1}+\gamma_{1}+\mu),-(\gamma_{2}+\mu+\delta_{2}) , -(\mu+\delta_{3}-\eta) \Bigg\}.$$
We see that the drug-free equilibria will be stable if $(\mu+\delta_{3})>\eta.$ Biologically this describes a situation where the sum of the natural death rate and removal rate (``smarten up" rate) is larger than the rate in which people complete the medical treatment. The reason why if there are more people completing treatement effects the stability is because when the previous drug-users complete treatment they get classified into another susceptible class where they may fall back into drug use.

For positive initial conditions, a plot illustrating the stability of the drug-free equilibrium is illustrated below.

\begin{figure}[!h]
  \centering
  \includegraphics[scale=0.52]{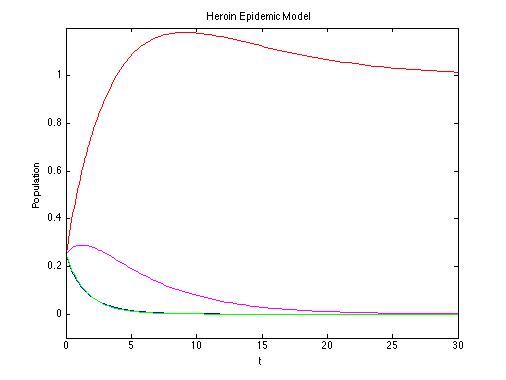}
  \footnotesize{\caption{Stability of the drug-free equilibria. Stability of this equilibrium depends on the sum of the natural death and removal rate to be greater than the rate in which people complete the medical treatment. This seemingly logical fallacy is because when a user completes treatment, they once again get put into a drug-free, but susceptible class.}}
  \label{WhiteAndComiskeyBoxes}
\end{figure}

For realistic purposes, it seems that this equilibria will be stable. This is due to the unfortunate fact of high relapse rates associated with most illicit drugs, which causes the probability for someone completing medical treatment to be small.  However, what is unsettling about the model is that it predicts if a devised treatment methodology is very successful, e.g., keeps the relapse rate minimal, then this drug-free equilibria will be unstable and cause the existence of a stable endemic steady-state. However, this is an artifact of the model. When drug-users successfully complete treatment and are put into a drug-free class in this model, they enter the susceptible class, who are equally likely to begin use.

%
%

\subsubsection{Endemic Steady-State}

When
\begin{equation}
\label{endemic_cond} (\mu+\delta_{3})<\eta,
\end{equation}
there will exist an endemic equilibria. We will solve for this solution numerically, and illustrate its existence through phase-plane analysis.

When using parameters values such that (\ref{endemic_cond}) is satisfied, we obtain a 3D-phase plane plot when plotting $s\ vs.\ l\ vs.\ h$, as seen in Figure(6).

\begin{figure}[!h]
  \centering
  \includegraphics[scale=0.5]{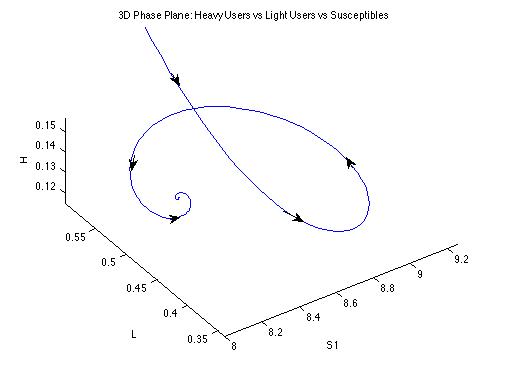}
  \footnotesize{\caption{Existence of Endemic Equilibria in the new drug model. The $3D$ phase plot compared the population of Heavy Users to Light Users and the Susceptible (drug-free) class. }}
  \label{Endemic_3d phase}
\end{figure}

%
%

%
%

\subsection{Results and Implications}

In the newly proposed model, we found that there exist two equilibrium, as in the White and Comiskey model, a drug-free equilibrium and an endemic steady-state. The drug-free equilibrium is found to be stable when the sum of the probabilities of natural death and enhanced removal rates is higher than the probability of someone successfully completing treatment. The model predicts that an endemic equilibria will arise if the probability of someone successfully completing treatment is increased. If a very successful treatment method is devised, then our model says that the drug-free equilibria will be unstable.

Moreover, the model does not predict that successful treatment is always optimal for eradicating drug-use. There are multiple ways to view this result. The first, as briefly mentioned in Section(\ref{stability_battista_model}), is that this purely an artifact of the model. It is possible the drug-users who successfully complete treatment should be put into a different drug-free class. One possibility would be to model this population to be completely drug-free, with no straight path to beginning the use the drug. However, this class may be treated as an intermediate step, in which a percentage of users must fall back into the susceptible class, and then may begin using the drug only after joining the susceptible population. 

The other way to view this result would be that successful treatment may be influencing people to begin using the drug, e.g., if an extremely successful and safe treatment exists, people may be less fearful to start using a drug. One deterrent of illicit drug-use is the non-existence of safe and successful rehabilitation and treatment programs. 

Our model also proposes a very apparent deterrent effect of heavy usage within a community, by introducing a new exponentially decreasing term on the traditional ``mass-action" term within Eqs.(\ref{BattistaModelS1}) and (\ref{BattistaModelL}). This term attempts to couple the effect of having a lot of heavy users in a community compared to light users, where if a large population of heavy users exist, the susceptible-1 class is much less likely to start using the drug since the dangers of addiction and usage are more apparent. The mathematical relationship that couple this type of behavior is an item to be considered and explored in a much deeper fashion.

Furthermore this this model can be applied to any new drug in which is highly addictive and hazardous in society. The existence of two drug-user classes reflects the reality of the situation, although the transience between these two populations needs more attention. In that vein, the flux of drug-users into treatment from both of these populations, is different and requires different strategies and treatments to successfully decrease both of these populations.

%
%

\section{Conclusion and Future Work}

We compare two different models of drug epidemics- White and Comiskey's model and our model, which includes two drug-user classes, a light-drug-user and heavy-drug-user class, a new susceptible class, which is comprised of post-successfully treated drug-users, and the unfortunate reality of successfully treated drug-users, who have finished their treatment and are now drug-free, but can relapse back into drug-use.  We find that in both models depending on values of parameters, there exist either a stable drug-free equilibria or a stable endemic equilibria.

For future work we wish to collect data to use in determining values of each parameter in the model, or in the very least estimating better values. We also need to find data to support whether or not we need the susceptible-2 class, or more specifically the values of $\beta_{1}$ and $\beta_{3}$. Furthermore we wish to construct a PDE dissipative model describing the spread of a drug from one more more singular points, such as a large city or a drug-den, onto the outlying community.

%
%

\section{Acknowledgements} 

The author would like to acknowledge Dr. Patricia Clark, whose introduction to mathematical biology course gave the motivation for this project, and Drs. David Ross and Anthony Harkin from R.I.T. for their insightful comments on mathematical model construction and the philosophy thereof.



\bibliographystyle{spmpsci}      
\bibliography{heroin}   

\end{document}